\documentclass[pre,12pt,preprint,showpacs,showkeys,amsmath]{revtex4}
\usepackage{eufrak}
\renewcommand{\vec}[1]{\mathbf{#1}}
\newcommand{\vecs}[1]{\boldsymbol{#1}}

\begin{document}

\noindent\fbox{
  \parbox{\textwidth}{
    Copyright notice: This article may be downloaded for personal use
    only. Any other use requires prior permission of the author and
    the American Physical Society. 

    This article appeared in "Physical Review E 67, 046310 (2003)",
    and may be found at https://doi.org/10.1103/PhysRevE.67.046310
  }
}

\title{Boundary conditions for probability density function transport equations in fluid mechanics}

\author{Luis Vali\~{n}o} 
\author{Juan Hierro}
\affiliation{LITEC, Consejo Superior de Investigaciones
Cient\'{\i}ficas\\
Mar\'{\i}a de Luna 3, Zaragoza 50018, Spain}

\date{\today}

\begin{abstract}
The behavior of the probability density function (PDF) transport equation at the limits of the probability space is studied from the point of view of fluid mechanics. 
Different boundary conditions are considered depending on the nature of the variable considered (velocity, scalar, position). A study of the implications  of entrance and exit conditions is performed, showing that a new term should be added to the PDF transport equation to preserve normalization in some nonstationary processes. In practice, this term is taken into account naturally in particle methods. Finally, the existence of discontinuities at the limits is also investigated.
\end{abstract}

\pacs{47.70.Fw,02.50.Ng,47.10.+g}
\keywords{Probability Density Function, Monte Carlo, Turbulent Combustion}

\maketitle

\section*{Introduction}
Methodologies based on the probability density functions (PDF)\cite{lundis67} of species are becoming increasingly popular for calculating turbulent reacting flows, due to the closed character of the chemical contribution to the PDF transport equation. After modeling the open terms in its transport equation, a Monte Carlo procedure is applied to numerically predict the PDF evolution\cite{popPDF85}.

The problem of the boundaries in stochastic processes has been considered in different books~\cite{garhan85,kartaysec81,risfok89}, mostly for the Fokker-Planck equation. For fluid mechanics related problems, there have been some works~\cite{klibilcon99, kuzsabtur90} which partially addresses this topic. In Pope's derivation, the case for a nonbounded magnitude was studied. 
As boundary conditions is an important issue in solving differential equations in fluid mechanics, a more comprehensive study has been carried out in this paper.

The Lagrangian frame will be considered as it offers more possible situations (the position is a stochastic variable), although the reasoning
is equally valid for an Eulerian one.

First, the boundary conditions in the case of a surface engulfing all the probability space is addressed. Sensible options are chosen depending on the variable and the specific problem considered. Next, the case of entrance and exit limits is studied. A new term in the transport equation will arise in the case of an imbalance between incoming and exiting particles. Finally the implications in modeling are considered and conclusions drawn. An appendix is added for the special case that Dirac deltas are accumulated on the boundary.
\section{Pdf transport equation for  boundaries covering the whole probability space}
There are different methods to derive PDF transport equations.
Probably the most popular is the one introduced by
Lundgren~\cite{lundis67}, which is based on Dirac's deltas
algebra. Another method which comes from a functional formulation was
developed by Dopazo and O'Brien~\cite{dopobrfun74}. We prefer here to use the method introduced by
Pope~\cite{popPDF85}, because the effect of the boundaries of the probabilistic space appears more explicitly.

In this section, we are going to focus on the general situations that commonly arise in fluid flows. Notice that
although the procedure is used for a PDF of a Lagrangian quantity  (nonconditional to the initial state, although this general case is added later), the reasoning is equally valid for an Eulerian magnitude.

Consider a test function (differentiable with continuous derivative in the domain) $Q \equiv
Q\left[\vec{c}^{+}(t)\right]$, where $\vec{c}^{+}(t)$ represents a
whole set of Lagrangian variables, possibly including chemical
species, position, velocity, etc. The evolution of its average,
including all variables, is next calculated,
\begin{equation}
 \frac{\mathrm{d}}{\mathrm{d}t}{\left\langle{Q}\right\rangle}  =
\frac{\mathrm{d}} {\mathrm{d}t} \int_V{Q(\vecs{\phi}) \,
{P_{\vec{c}^{+}} }(\vecs{\phi};t) \, d\vecs{\phi}} =   \int_s{Q \,
\frac{\mathrm{d}{\phi}^s_\alpha}{\mathrm{d}t} \, n^s_\alpha \,
{P_{\vec{c}^{+}} } \, ds} + \int_V{Q \,
{\frac{\mathrm{d}P_{\vec{c}^{+}}}{\mathrm{d}t}} \, d\vecs{\phi}}.
 \label{lPDF1}
\end{equation}
The first term on the right hand side of the previous equation,
which has been considered explicitly  (at difference with previous
works) is a consequence of the evolution with time of the limits
$\vecs{\phi}^s$ of the PDF. These limits form a surface $s$ which
bounds $V$, the accessible region in the probabilistic space. It has
outward unit vector $\vec{n}^s$.

We are considering that $s$ covers all the accessible values in the probabilistic space, so no particles are allowed to cross this boundary . The situation of an infinite domain has been studied by Pope \cite{popPDF85}. In this paper the $s$ is not necessarily at the infinite.

On the other hand, as taking averages and derivatives commute,
\begin{eqnarray}
\nonumber
\frac{\mathrm{d}}{\mathrm{d}t}{\left\langle{Q}\right\rangle}  & =
& \left\langle {\frac{\mathrm{d}}{\mathrm{d}t} Q}\right\rangle  =
\int_V{ \left\langle \left.
 \frac{\mathrm{d}}{\mathrm{d}t} Q \right|
\vecs{\phi}  \right\rangle \, P_{\vec{c}^{+}} \, d\vecs{\phi}} \\
 & = & \int_V{
\left\langle \left. \frac{\partial Q}{\partial c^+_\alpha}
\frac{\mathrm{d} c^+_\alpha}{\mathrm{d} t} \right|  \vecs{\phi}
\right\rangle P_{\vec{c}^{+}} \, d\vecs{\phi}} = \int_V{
\frac{\partial Q}{\partial \phi_\alpha} \left\langle \left.
 \frac{\mathrm{d}
c^+_\alpha}{\mathrm{d} t} \right|  \vecs{\phi} \right\rangle
P_{\vec{c}^{+}} \, d\vecs{\phi}}
 \label{lPDF2}
\end{eqnarray}
where the second equality comes from a well-known PDF property
regarding conditional averages~\cite{popPDF85}, the third one is a direct
application of the chain rule, and the last one comes from the
fact that $Q$ derivatives are uniquely functions of $\vec{c}^{+}$,
so they come out of the conditional average as functions of
$\vecs{\phi}$ .

Equation (\ref{lPDF2}) is next integrated by parts. Using the
divergence theorem to pass from a volume to a surface integral:
\begin{eqnarray}
\frac{\mathrm{d}}{\mathrm{d}t}{\left\langle{Q}\right\rangle}  & =
&  \int_s{  Q \left\langle \left. \frac{\mathrm{d}
c^+_\alpha}{\mathrm{d} t} \right|  \vecs{\phi}^s \right\rangle
 \, n^s_\alpha \,P_{\vec{c}^{+}} \, ds} - \int_V{ Q \frac{\partial}{\partial \phi_\alpha}\left(\left\langle
\left. \frac{\mathrm{d} c^+_\alpha}{\mathrm{d} t} \right|
\vecs{\phi} \right\rangle P_{\vec{c}^{+}}\right) d\vecs{\phi}}.
 \label{lPDF3}
\end{eqnarray}
Now, equalizing the integrals in the last terms in
Eqs.(\ref{lPDF1}) and (\ref{lPDF3}), grouping volume and surface integrals, the following equation is obtained:
\begin{eqnarray}
\int_V{Q \, \left[
{\frac{\mathrm{d}P_{\vec{c}^{+}}}{\mathrm{d}t}}  +  \frac{\partial}{\partial \phi_\alpha}\left(\left\langle
\left. \frac{\mathrm{d} c^+_\alpha}{\mathrm{d} t} \right|
\vecs{\phi} \right\rangle P_{\vec{c}^{+}} \right) \right]\, d\vecs{\phi}} & + &  \nonumber \\ 
 \int_s{Q \, \left(
\frac{\mathrm{d}{\phi}^s_\alpha}{\mathrm{d}t} \, n^s_\alpha \,
{P_{\vec{c}^{+}} } - \left\langle \left. \frac{\mathrm{d}
c^+_\alpha}{\mathrm{d} t} \right|  \vecs{\phi}^s \right\rangle
 \, n^s_\alpha \,P_{\vec{c}^{+}} \right)\, ds} & = &0. 
\label{lPDF4a}
\end{eqnarray}

Next, it is going to be assumed that ${P_{\vec{c}^{+}} }$ has no Dirac delta kind of singularities, which could convert volume integrals into surface integrals. That means that in situations like a scalar binary mixing $(0,1)$ problem, regions with the extremal values of the scalar $1$ or $0$ are not to be encountered with a noninfinitesimal measure. These singularities only arise in the limit of infinite Reynolds number, infinitely fast chemical reactions\cite{kuzsabtur90} or infinitely compressible fluid. For the sake of completeness, this special case is included in the Appendix.

With the previous assumption about ${P_{\vec{c}^{+}} }$ and taking into consideration that the smooth function $Q$ is arbitrary, the volume integrand and the surface integrand should be separately equal to zero. So, finally, the following equations are obtained:
\begin{eqnarray}
{\frac{\mathrm{d}P_{\vec{c}^{+}}}{\mathrm{d}t}}  +  \frac{\partial}{\partial \phi_\alpha}\left(\left\langle
\left. \frac{\mathrm{d} c^+_\alpha}{\mathrm{d} t} \right|
\vecs{\phi} \right\rangle P_{\vec{c}^{+}} \right) & = & 0  
\label{lPDF5}\\
j ^s _\alpha  \, n^s_\alpha \, \equiv \left\langle \left. \frac{\mathrm{d}
c^+_\alpha}{\mathrm{d} t} \right|  \vecs{\phi}^s \right\rangle
 \, n^s_\alpha \,P_{\vec{c}^{+}} - \frac{\mathrm{d}{\phi}^s_\alpha}{\mathrm{d}t} \, n^s_\alpha \,
{P_{\vec{c}^{+}} } & = &0,
\label{lPDF6}
\end{eqnarray}
where the definition of the probability current $\vec{j}$ is included.

Equation (\ref{lPDF5}) is the well-known result which describes the evolution of the PDF in the domain encircled by the surface $s$ in terms of the conditional averages of the temporal derivatives of its variables.

Equation (\ref{lPDF6}) arises from the fact that we are explicitly considering that surface and gives the rule for the evolution of $s$ in terms of the PDF and the same conditional averages evaluated on the surface. This surface engulfs the whole probability space as it is implicitly assumed when writing the first equality in Eq. (\ref{lPDF1}). Equation (\ref{lPDF6}) guarantees that no flux of particles crosses $s$, so $s$ will keep containing the whole probability space in the future.

Notice that there are no fluctuations of $\frac{\mathrm{d}c^+_\alpha}{\mathrm{d} t}$ about its conditional mean (except for, at most, a measure nil set), since a positive value (when projecting to the normal $\vec{n}$)) of this quantity would violate 
boundedness. To understand this restriction, it is useful to write the following expression of the net flux of probability between two regions $V_2$ and $V_1$ as a function of two-time probability density functions\ \cite{garhan85}
\begin{equation}
\lim_{\Delta t \to 0} \frac{1}{\Delta t} \int_{V_1 }\mathrm{d}{\vecs{\phi}_1}\int_{V_2 }\mathrm{d}{\vecs{\phi}_2}\left[ P\left(\vecs{\phi}_1; t+\Delta t ,\vecs{\phi}_2; t\right) - P\left(\vecs{\phi}_2; t+\Delta t , \vecs{\phi}_1; t\right)\right] = \int_{s_{12} }\mathrm{d}s \; j ^{s_{12}}_\alpha   n^{s_{12}}_\alpha,
\label{gar1}
\end{equation}
where $P\left(\vecs{\phi}_1; t+\Delta t ,\vecs{\phi}_2; t\right)$ is the probability density of the event $$\left\{ \vec{c}^+(t)=\vecs{\phi}_2 \: \mathit{and} \: \vec{c}^+(t+~\Delta t)=\vecs{\phi}_1 \right\},$$ $s_{12}$ is the surface separating both regions (for simplicity, assumed constant), and the normal $\vec{n}^{s_{12}}$ points from $V_2$ to $V_1$. In Eq. (\ref{gar1}), it is also assumed, as in the rest of this section, that there are no Dirac delta contributions.
 
In the present situation, as there are no particles crossing the surface $s \equiv s_{12}$, both two-point probabilities are zero ($V_2$ is our probabilistic space region $V$). That means that $P\left(\vecs{\phi}_1; t+\Delta t ,\vecs{\phi}_2; t\right)=0$ for any point in $V \equiv V_2$  and $\vecs{\phi}_1$ representing any point outside $V$.  In particular, in the event of a particle very close to $s$, that probability is also zero, which implies that it should have its $\frac{\mathrm{d}c^+_\alpha}{\mathrm{d} t}$ ``randomness'' diminished in order not to cross the surface. In the limit, on $s$, the conditional variance of
$\frac{\mathrm{d}c^+_\alpha}{\mathrm{d} t}$ should be zero. Obviously, as there are no particles outside $V_2$, the other PDF $P\left(\vecs{\phi}_2; t+\Delta t ,\vecs{\phi}_1; t\right)$ is zero too.

It is noticed that this is a consequence of the forbidden crossing restriction, which is, in a fluid mechanics application, a physical constraint of some type. For example if $\vec{c}^+$ is the fluid particle position, a wall which prevents that crossing would be the boundary $s$. In the following section, the noncrossing restriction is removed and consequences derived. Also notice that in the case of $s$ at infinite, as long as  $\left\langle Q \frac{\mathrm{d}
c^+_\alpha}{\mathrm{d} t}  \right\rangle$ exists, the modulus of the integrand of Eq. (\ref{lPDF4a}) should be null~\cite{popPDF85}.

There are two essential ways of satisfying the condition expressed by Eq. (\ref{lPDF6}). The first one is to keep $s$ ``far away'' enough as to have null probability of having particles on the surface; that is $P_{\vec{c}^{+}}(\phi^s)=0$. The movement of the points which constitute the surface is irrelevant (as long as $P_{\vec{c}^{+}}(\phi^s)=0$ is guaranteed).  This case could be appropriate in a situation where  $\vec{c}^{+}$ is a velocity field whose values are known not to reach certain limits.  The second one is allowing probability density of having particles on the surface, but keeping the probability current at any point of the boundary $s$, $j ^s _\alpha  \, n^s_\alpha$, null. This is appropriate, for example, for the $(0,1)$ binary mixing problem (finite Reynolds number). $0$ and $1$ form the initial boundary of the probability space.  In this case, it is possible either to keep the boundaries fixed (as after the initial state, the probability of having the extreme values will be nil) or  move $s$ at the same (or smaller) rate as the conditional average. Moving $s$ at the same rate as the conditional average at all times means that
\begin{equation}
\frac{\mathrm{d}{\phi}^s_\alpha}{\mathrm{d}t} \, n^s_\alpha 
= \left\langle \left. \frac{\mathrm{d}
c^+_\alpha}{\mathrm{d} t} \right|  \vecs{\phi}^s \right\rangle
 \, n^s_\alpha 
\label{lPDF7}
\end{equation}
This way, if  $s(\vecs{\phi})=0$  is initially the minimum surface surrounding the whole of the accessible probabilistic space, it would remain this way for all times.

As the movements of  $\vecs{\phi}^s$ parallel to $s$ are irrelevant in order to define the surface, there are $n-1$ degrees of freedom in defining its evolution in Eq. (\ref{lPDF7}). Arbitrarily, and for convenience, the simple choice
 \begin{equation}
\frac{\mathrm{d}{\phi}^s_\alpha}{\mathrm{d}t}
 = \left\langle \left. \frac{\mathrm{d}
c^+_\alpha}{\mathrm{d} t} \right|  \vecs{\phi}^s \right\rangle 
 \label{lPDFss}
\end{equation}
is taken. Equation (\ref{lPDFss}) determines a natural evolution of the PDF boundaries. It
shows that the PDF limits can be chosen to evolve in a natural way following the same rules as the
notional particles representing the PDF.

The choice of this particular boundary is also appropriate in situations when special care in the application of the derivation rules (the chain rule or
integration by parts) is needed. These rules demand
continuous derivatives for the PDF in the interior domain of the
function. It is not clear that the PDF of a bounded scalar, as a
chemical species, does not have a sudden drop in its limiting
value. If this is so, the derivative would be discontinuous
(although both left and right derivatives exist). The previous
deduction would be still valid as the left derivative exists on the
bound, but it would not be valid a deduction which keeps the
original limits fixed (of moving with inappropriate rate), so the point with the sudden drop would be interior.

Going back to Eq. (\ref{lPDF4a}), any allowable evolution of the boundary surface implies that the second term in the right-hand side vanishes, so this equation can be simplified as follows:
 \begin{equation}
{\frac{\mathrm{d}P_{\vec{c}^{+}}}{\mathrm{d}t}} = -
\frac{\partial}{\partial \phi_\alpha} \left( \left\langle \left.
\frac{\mathrm{d} c^+_\alpha}{\mathrm{d} t} \right| \vecs{\phi}
\right\rangle P_{\vec{c}^{+}} \right) 
\label{lPDFv}
\end{equation}
This is the well-known result which provides the evolution rule for a PDF transport equation from the conditional averages of temporal derivatives of its variables in physical space.

Finally, a consideration about conditioning on constant quantities. Consider an additional set of variables in the PDF which are constant in time. We use the notation with subindex ``zero'' for this extra set. If the PDF $P_{\vec{c}^{+}\vec{c}_0^{+}}$ is decomposed  as $P_{\left. \vec{c}^{+}\right| \vec{c}_0^{+}}P_{\vec{c}_0^{+}}$, direct substitution into Eq. (\ref{lPDFv}), and taking into account that any average or PDF of the constant variables does not change with time, it is trivially obtained:
\begin{equation}
{\frac{\mathrm{d}P_{\left. \vec{c}^{+}\right| \vec{c}_0^{+}}}{\mathrm{d}t}} = -
\frac{\partial}{\partial \phi_\alpha} \left( \left\langle \left.
\frac{\mathrm{d} c^+_\alpha}{\mathrm{d} t} \right| \vecs{\phi} , \vecs{\phi}_0
\right\rangle P_{\left. \vec{c}^{+}\right| \vec{c}_0^{+}} \right).
\label{lPDFvc}
\end{equation}
Equation (\ref{lPDFvc}) was first obtained by Pope in Ref.~\cite{popPDF85} using Lundgren's methodology. Notice that a similar result for the Fokker-Planck equation is shown by Gardiner~\cite{garhan85}.

By the application of the same idea to the evolution for the moving PDF limits, it is also trivial:
\begin{equation}
\frac{\mathrm{d}{\phi}^s_\alpha}{\mathrm{d}t}
 = \left\langle \left. \frac{\mathrm{d}
c^+_\alpha}{\mathrm{d} t} \right|  \vecs{\phi}^s ,\vecs{\phi}_0^s \right\rangle .
 \label{lPDFssc}
\end{equation}
\section{Pdf transport equation for  boundaries not covering the whole probability space}
The mathematical development in the preceding section does not provide  for situations with flux of probability through the boundaries. That was necessary in order to preserve the normalization of the PDF. However, it is  possible to have situations where there is a flux of particles through the boundaries, which implies a flux of probability leading, in principle, to an improperly normalized PDF. This is the case, for example, when the stochastic variable is the position and in a nontransient situation, more fluid particles leave than enter the domain per unit time. The apparent contradiction lies in the fact that the PDF that is actually simulated in such a case is the PDF of the position conditional on having the values inside the domain. For that reason the formulation derived in this section is also appropriate for the case that the net probability current is nil on each point of the limiting surface, $s$, but particles can cross the boundary, because of the existence of particles outside our probabilistic domain. In this situation, even if the conditional mean of $\frac{\mathrm{d}c^+_\alpha}{\mathrm{d} t}$ is zero, it may have a variance, as long as $P\left(\vecs{\phi}_1; t+\Delta t ,\vecs{\phi}_2; t\right) = P\left(\vecs{\phi}_2; t+\Delta t , \vecs{\phi}_1; t\right)$ in Eq. (\ref{gar1}). 

The proper way of obtaining the transport equation of such PDFs is to consider first the equation of the PDF over all the space, and then to apply the relationship that exists between them. This is done next.

Consider $V'$ the domain of  the probabilistic  space, we are interested in. This region is encircled by a surface $s'$. For example, if we are thinking of $\vec{c}^{+}=\vec{x}^{+}$ for a turbulent jet, $s'$ could be formed by the entry, exit and suitable lateral limits.  We consider  $V'$  contained inside the whole probabilistic space $V$, which is in its turn embraced by a surface $s$. If $s'$ shares some part with $s$, the nonflux conditions explained in the preceding section would apply to this part, so for simplifying the analysis, this situation will not be considered.  Eq. (\ref{lPDFv}) would be then fulfilled in $V'$  and also in $s'$. Then it is possible to integrate that equation, multiplied by a test function $Q$ satisfying the same properties as the one in the section I:
\begin{equation}
\int_{V'}{Q \, \left[
{\frac{\mathrm{d}P_{\vec{c}^{+}}}{\mathrm{d}t}}  +  \frac{\partial}{\partial \phi_\alpha}\left(\left\langle
\left. \frac{\mathrm{d} c^+_\alpha}{\mathrm{d} t} \right|
\vecs{\phi} \right\rangle P_{\vec{c}^{+}} \right) \right]\, d\vecs{\phi}}= 0,
\label{PDFacotadaori}
\end{equation}
where $V'$ at the bottom of the integral sign express that the integration is performed in that volume. 

Now $P_{\vec{c}^{+}}$ is substituted to obtain the transport equation of $P_{ \left. \vec{c}^{+}  \right| V'}$, the PDF conditional on  the values of $\vec{c}^{+}$ being in $V'$. This is done by the existing following relationship between both PDFs,
\begin{equation}
P_{ \left. \vec{c}^{+}  \right| V'}=\frac{P_{\vec{c}^{+}}}{\aleph},
\label{relacionPDFs}
\end{equation}
where
\begin{equation}
\aleph=\int_{V'}{P_{\vec{c}^{+}} \, d\vecs{\phi}}
\label{normalizacion}
\end{equation}
guarantees the normalization of $P_{ \left. \vec{c}^{+}  \right| V'}$. The result of this substitution is
\begin{equation}
\int_{V'}{Q \, \left[
{\frac{\mathrm{d}P_{ \left. \vec{c}^{+}  \right| V'}}{\mathrm{d}t}} \, \aleph + P_{ \left. \vec{c}^{+}  \right| V'} \, \dot{\aleph} + \aleph \, \frac{\partial}{\partial \phi_\alpha}\left(\left\langle
\left. \frac{\mathrm{d} c^+_\alpha}{\mathrm{d} t} \right|
\vecs{\phi} \right\rangle P_{ \left. \vec{c}^{+}  \right| V'} \right) \right]\, d\vecs{\phi}}= 0. 
\label{PDFacotada}
\end{equation}
Being $Q$ a general enough test function and by the same arguments as in the preceding section, the following differential form equivalent to  Eq. (\ref{lPDFv}) is obtained:
 \begin{equation}
{\frac{\mathrm{d}P_{ \left. \vec{c}^{+}  \right| V'}}{\mathrm{d}t}} + P_{ \left. \vec{c}^{+}  \right| V'} \, \frac{\dot{\aleph} }{\aleph}+  \frac{\partial}{\partial \phi_\alpha}\left(\left\langle
\left. \frac{\mathrm{d} c^+_\alpha}{\mathrm{d} t} \right|
\vecs{\phi} \right\rangle P_{ \left. \vec{c}^{+}  \right| V'} \right) = 0.
\label{PDFacotada2}
\end{equation}
The first and third terms in  Eq. (\ref{PDFacotada2}) appear also in Eq. (\ref{lPDFv}), and have the same meaning. The second one is the consequence of the flux of probability through the boundaries and should be interpreted as a renormalization of the PDF.  In fact, (minus) $\aleph$  time derivative $\dot{\aleph}$ is the net flux of probability of $P_{\vec{c}^{+}}$ through $s'$, $J^{s'}$, as it is readily shown from taking the time derivative of Eq. (\ref{normalizacion}), replacing the resulting PDF time derivative by means of Eq. (\ref{lPDFv}) and using the same mathematics as in Eq. (\ref{lPDF1}):
\begin{equation}
 \dot{\aleph}=\int_{s'}{\left(
\frac{\mathrm{d}{\phi}^{s'}_\alpha}{\mathrm{d}t} \, n^{s'}_\alpha \,
{P_{\vec{c}^{+}} } - \left\langle \left. \frac{\mathrm{d}c^+_\alpha}{\mathrm{d} t} \right|  \vecs{\phi}^{s'} \right\rangle
 \, n^{s'}_\alpha \,P_{\vec{c}^{+}} \right)\, d\vecs{\phi}^{s'}} = - \int_{s'}{{j}_\alpha^{s'} n_\alpha^{s'}\, d\vecs{\phi}^{s'}}= - J^{s'}.
 \label{normalizacionpunto}
\end{equation}
Another way of looking at this renormalization is by expressing the PDF  in terms of Dirac delta functions, as it is done in Monte Carlo PDF methods:
\begin{equation}
P_{ \left. \vec{c}^{+}  \right| V'} = \frac{1}{N} \sum_{n=1}^{N}  \delta\left[\vecs{\phi}-{\vec{c^+}}^{n}_{\left .  \right| V'}(t)\right] .
\label{deltaPDF}
\end{equation}
If now $N$ is allowed to change in time and the value of the particles $\vec{c^+}$ kept constant, the following transport equation is obtained:
\begin{equation}
{\frac{\mathrm{d}P_{ \left. \vec{c}^{+}  \right| V'}}{\mathrm{d}t}} = -\frac{\dot{N}}{N^2} \sum_{n=1}^{N}  \delta\left[\vecs{\phi}-{\vec{c^+}}^{n}_{\left .  \right| V'}(t)\right]=-\frac{\dot{N}}{N} P_{ \left. \vec{c}^{+}  \right| V'} ,
\label{deltaPDF2}
\end{equation}
where $N$ is big enough and the temporal step is small enough as to allow the notation abuse $\dot{N}$ . 

It is convenient to express the flux in Eq. (\ref{PDFacotada2}) in terms of conditional PDF. This is done by applying Eq. (\ref{relacionPDFs}) to Eq. (\ref{normalizacionpunto}), obtaining
\begin{equation}
 \frac{\dot{\aleph}}{\aleph}=\int_{s'}{\left(
\frac{\mathrm{d}{\phi}^{s'}_\alpha}{\mathrm{d}t} \, n^{s'}_\alpha \,
{P_{ \left. \vec{c}^{+}  \right| V'} } - \left\langle \left. \frac{\mathrm{d}c^+_\alpha}{\mathrm{d} t} \right|  \vecs{\phi}^{s'} \right\rangle
 \, n^{s'}_\alpha \,P_{ \left. \vec{c}^{+}  \right| V'} \right)\, d\vecs{\phi}^{s'}} = - \int_{s'}{{{j}_\alpha^{s'}}_{\left .  \right| V'} n_\alpha^{s'}\, d\vecs{\phi}^{s'}}= - J^{s'}_{\left .  \right| V'}
 \label{normalizacionpunto2}
\end{equation}
in obvious notation.

With this result, the final form of the transport equation for the conditional PDF is 
 \begin{equation}
{\frac{\mathrm{d}P_{ \left. \vec{c}^{+}  \right| V'}}{\mathrm{d}t}} - J^{s'}_{\left .  \right| V'} P_{ \left. \vec{c}^{+}  \right| V'} +  \frac{\partial}{\partial \phi_\alpha}\left(\left\langle
\left. \frac{\mathrm{d} c^+_\alpha}{\mathrm{d} t} \right|
\vecs{\phi} \right\rangle P_{ \left. \vec{c}^{+}  \right| V'} \right) = 0,
\label{PDFacotada2bis}
\end{equation}
with the boundary conditions given by $\vec{j}^{s'}_{\left .  \right| V'} \cdot \vec{n}^{s'}$, the outgoing current of probability normal to each point of $s'$.

When the net flux is null, the second term disappears and no renormalization is required. This situation happens, as explained, at the beginning of this section, when the probability density which is transported by entering particles compensates the one removed by exiting ones. But in transient situations, there is a contribution, which proportionally increases or decreases $P_{ \left. \vec{c}^{+}  \right| V'}$, depending on the outward or inward character of the net flux of probability $J^{s'}_{\left .  \right| V'}$. The proper normalization of  $ P_{ \left. \vec{c}^{+}  \right| V'}$ is always guaranteed.

\section{Consequences in the numerical implementation of PDF methods}
Monte Carlo methods, used to numerically solve the PDF transport equation, represent that PDF by a set of stochastic particles. The evolution of these particles should be in such a way that their one point PDF of the considered magnitudes is close enough (ideally identical) to that of the real flow. But each stochastic particle does not have to behave as a fluid particle. In fact, the length and time scales used for the numerical algorithms in particle methods are much bigger than these required by direct numerical simulations. In general, their evolution is represented by all kind of Markovian processes, including nondifferentiable, as Ornstein-Uhlenbeck, and even noncontinuous evolutions, as jump processes. The current of probability should include obviously the contribution of all these kind of stochastic processes.

In this section, the consequences of the previous results in the application of PDF methods to turbulent flows are studied. It will be seen that these consequences are in practice taken into account naturally in particle methods. Nevertheless, it is enlightening to justify these practices, understanding and taking advantage of the close relationship between particle and PDF evolutions; particularly, in the case of  a probability current crossing the boundaries of the probabilistic domain.

The proper way of applying boundary conditions in PDF methods is by studying $ \vec{j}^{s} \, \vec{n}^{s}$ for the PDF expressed in terms of the real fluid particles. For reasonable choices of the PDF boundary limits, this quantity should be deduced in terms of the boundary conditions of the transport equation of the magnitude studied. Then, the same quantity should have its value reproduced by the modeling stochastic particles. That means that first the boundary conditions of the real field (assuming they are coming from the Navier-Stokes equations) are considered. From these, proper boundary conditions for the PDF of that real field are deduced. And finally, these ones will be the guide for establishing the boundary conditions of the PDF of the modelled field. It is reminded that the aim of PDF modeling is that the PDF of the modelled and real field are as closed as possible.  Examples are next given.

\subsubsection{No crossing through the boundary allowed.}  
As real fluid particles are not allowed to cross the boundary in this case, the particles representing the PDF cannot cross the boundary either, whatever the stochastic process is chosen to model their behavior. The specific way of fulfilling this condition depends on the physics of the magnitude being solved and, ultimately, on the modeling chosen for the stochastic process to represent this physics. 

In the case of real fields with nonbounded magnitudes, as the velocity, all the possible choices are expressed, in practice, by nonrestrictions in the values that $\vec{c}^{+}$ can reach. That is to say, the velocity values which a particle can reach due to a stochastic process are not restricted in any way.

In the case of a real bounded magnitude, as a scalar mass fraction, the stochastic processes used to model the PDF should avoid that particles cross the limiting values, as real particles do. From a practical point of view, some rule is derived to numerically avoid this crossing. As mentioned in section I, this approach is used in binary mixing problems for scalars. In fact, many mixing models guarantee boundedness by construction.
\subsubsection{Crossing through the boundary allowed.}
The only magnitude, of those that appear in a flow, which can reasonably be affected by this kind of boundary condition is the position.  In this case, the current of probability through the entry and/or exiting boundary $s'$ should be given. In view of Eq. (\ref{lPDF6}), this is the conditional velocity at $s'$.  This requirement is fulfilled  by knowing the velocity and number of particles which enter and leave the domain. In transient situations, it is also necessary to implement the renormalization term in Eq. (\ref{PDFacotada2}).  This is done by reevaluating $N$ counting the particles remaining in the domain in each time-step as it is immediate from Eqs. (\ref{deltaPDF}) and (\ref{deltaPDF2}). In fact, in practical implementation of particle methods, the mass density function $\cal F$ is used, which is defined as the total mass in the considered volume times the (density weighted) Lagrangian PDF. It is immediate to prove when taking the time derivative of  $\cal F$, that the term coming from the variation of the total mass cancels with the renormalization term shown in Eq. (\ref{deltaPDF2}), so the same equation is obtained for $\cal F$ regardless of the volume considered. In any case, from the previous analysis, it is clear that the (density weighted) Lagrangian PDF to be used in the definition of $\cal F$ should be conditional to the particles being inside the domain $V'$.
\subsection{Eulerian Frame}
Although the results shown for the Lagrangian frame are equally valid in the Eulerian one, there are some specific considerations that affect the boundary conditions of the PDF in the Eulerian frame. Given the introduction of numerical methods which use Monte Carlo fields instead of particles\cite{valfie98}, it is convenient to point them out.

The fundamental difference between the Eulerian and Lagrangian frames, from the mathematical point of view, is the different role that position $\vec{x}$ plays.  In the Lagrangian frame, $\vec{x}$ is a variable in the probabilistic sense, while in the Eulerian one, $\vec{x}$ is a variable in the ordinary sense. As a consequence, boundary conditions related with $\vec{x}$ should be treated in the ordinary function sense. For example, stochastic fields representing temperature should be consistent with the boundary condition of the PDF, as an ordinary function of the position, on the wall.  Let's say the PDF of temperature on the wall is given to be a Gaussian. Then the stochastic fields will have Gaussian distributed values on the wall. Other Dirichlet, Neumann or mixed boundary conditions should be similarly applied.

Notice that the meaning of the $\frac{\mathrm{d}}{\mathrm{d}t}$ operator in Eq. (\ref{lPDFv}) and (\ref{PDFacotada2bis}) depends on the character of the frame of reference used to represent the PDF. In a Lagrangian frame, it is equivalent to $\frac{\partial}{\partial t}$, with the conditional mean velocity as a flux in the probabilistic space. In an Eulerian frame of reference, $\frac{\mathrm{d}}{\mathrm{d}t}$ has the usual meaning of $\frac{\partial}{\partial t} + u_i \frac{\partial}{\partial x_i}$, with no random variables associated to position.
\section{Conclusions}
The behavior of the PDF transport equation 
at the limits of the probability space has been  studied, considering 
the variables that usually appear in fluid mechanics. In all cases, the important quantity to be given at the boundary is the current of probability. Different boundary conditions are contemplated depending on the nature of the variables considered; velocity, scalars and position. In the case of unbounded magnitudes, as the velocity, the boundary is at the infinity (case already studied by Pope\cite{popPDF85}). 
In the case of bounded magnitudes, as mass fractions, there are two theoretically possible cases: 
a fix boundary in the limit of nonaccessible values or a boundary 
which moves with the ``velocity'' conditional on the fluid particle 
being on that boundary. Finally, for the position of the fluid particles, 
it is shown that the actual PDF studied is the PDF of the position 
conditional on the particle being confined in the spatial domain considered 
in the problem. The existence of a new term is proved in order to keep 
the PDF normalized in some transient situations. 

Although much of these results are implicitly taken into account in particle methods, it is enlightening to explicitly justify these practices. Notice that the close relationship between particle and PDF evolutions allows the fully understanding of special situations like the case of a probability current crossing the boundaries of the probabilistic domain.

For the sake of completeness, the limiting cases, when there are 
discontinuities at the limits, have been also investigated and shown 
in the Appendix.
\section{Appendix}
In this appendix, the PDF is allowed to have possible singularities on the boundaries in the form of Dirac delta contributions. As mentioned above, these singularities only arise in the limit of infinite Reynolds number or in the limit of infinitely fast chemical reactions (or in other more unusual limits, as in the case of a infinitely compressible flow or infinite Mach number).  Kuznetsov and Sabelnikov~\cite{kuzsabtur90} studied external intermittence in single scalar PDFs with this kind of singularity.  They consider the PDF of a single scalar through a turbulent  mixing layer, including the upper ($c=1$) and down ($c=0$) nonturbulent regions.

If the Dirac delta singularity arose in the inner domain, it would be always possible to split it using the singularities as boundaries, although it is not certainly a common situation in turbulent reacting flows.

First, it is convenient to express $P_{\vec{c}^{+}}$ as a regular part $P^r_{\vec{c}^{+}}$ plus the possible Dirac delta contribution at the boundary:
\begin{equation}
 P_{\vec{c}^{+}}= P^r_{\vec{c}^{+}} + \gamma \, \delta_s 
 \label{PDFescom}
\end{equation}
where $\gamma(\vecs{\phi}, t)$ indicates the weight of the Dirac delta contribution $\delta_s $ to the total PDF.  It is defined as a function $\gamma_s(\vecs{\phi}^s, t)$  on the boundary and takes arbitrary values in the rest of $V$.

More properly, the generalized function $\delta_s$, usually called "simple layer"  is a multidimensional  extension of the Dirac delta which transform volume integrals into surface integrals on the surface $s$. In the case of a spherical surface and in spherical coordinates, $\delta_s= \delta\left(r-r^s \right)$  and it is known as the radial Dirac delta function. Notice that $\gamma_s \ge 0$ cannot be arbitrarily big as the normalizing condition $\int_V{
{P_{\vec{c}^{+}} }(\vecs{\phi};t) \, d\vecs{\phi}} = 1$ should be preserved.

Kuznetsov and Sabelnikov follow a different  approach for the one-dimensional case; the discontinuity is fixed and they multiply $P^r_{\vec{c}^{+}}$  by a generalized function that takes the value 1 inside $s$ and  0 outside, and integrate in an infinite domain.

Now the procedure used in section I is repeated here. The contribution of the regular part of the PDF is given by Eq. (\ref{lPDF4a}), replacing  $P_{\vec{c}^{+}}$ by $P^r_{\vec{c}^{+}} $ . The contribution of the discontinuity is next deduced. Considering again the test function $Q$, with the same properties as in section I, the analogous to Eq. (\ref{lPDF1}) is
\begin{eqnarray}
 & &\frac{\mathrm{d}} {\mathrm{d}t} \int_V{Q(\vecs{\phi}) \,
{\gamma( \vecs{\phi})\, \delta_s} \, d\vecs{\phi}} \overset{1}{=}  \frac{\mathrm{d}} {\mathrm{d}t}\int_s{Q(\vecs{\phi}^s) \,
{\gamma_s( \vecs{\phi}^s)  } \, ds} \overset{2}{=}  \frac{\mathrm{d}} {\mathrm{d}t}\int_s{Q(\vecs{\psi}^s) \, {\gamma_s( \vecs{\psi}^s)   } \, g ^{\frac{1}{2}}\, d\vecs{\psi}^s} 
= \nonumber \\
 &   &\overset{3}{=}\int_s{  \left\{ \frac{\mathrm{d}} {\mathrm{d}t}
 \left[  Q(\vecs{\psi}^s) \, \gamma_s( \vecs{\psi}^s) \right] + Q(\vecs{\psi}^s) \, \gamma_s( \vecs{\psi}^s)\,  g ^{-\frac{1}{2}}\frac{\partial } {\partial \psi^{j} }\left( g ^{\frac{1}{2}}\frac{\mathrm{d}\psi^{sj}}  {\mathrm{d}t}\right)  + Q(\vecs{\psi}^s) \, \gamma_s( \vecs{\psi}^s)   \frac{1}{2 g}\frac{\mathrm{d}g} {\mathrm{d}t}\right\}\, g ^{\frac{1}{2}}\, d\vecs{\psi}^s}, \nonumber \\
\label{lPDF1ext}
\end{eqnarray} 
where equality 1 comes from the definition of $\delta_s$; equality 2 reflects a coordinate change to $\vecs{\psi}$, with $\psi_j, \{j=2,...,N\}$ being generalized surface coordinates (denoted globally as $\vecs{\psi}^s$), $\psi^1$ being the normal-to-the surface coordinate and $g$ the determinant of the metric tensor ($g^{\frac{1}{2}}$ is the Jacobian); and equality 3 is the Reynolds transport theorem for surfaces. See, for example, the book by Aris\ \cite{vecari90} for the mathematical details. 

It has been applied a new convention for repeated indices: Latin go from 2 to $N$ (surface) and Greek from 1 to $N$ . Contravariant components used. Notice that the determinant of the metric tensor $g$ is the same for surface and volume coordinates $\boldsymbol{\psi}$, as $g_{1\alpha}=\delta_{1\alpha}$, which comes from the fact that the first coordinate is normal to the others and its correspondent vector has unity length. Different symbols should have used to express the different functional dependency of $Q$ and $\gamma_s$ on the new system coordinate, but they have been written the same in order to not overload the notation. The context will be used to distinguish instead.

Notice that  $c^{\alpha}$ is the contravariant component $\alpha$  in the $\vecs{\psi}$ coordinate system, which is not  the actual scalar $\alpha$ in the original $\vecs{\phi}$ system, although the same symbol has been used, as for $Q$ and $\gamma_s$. 

It is convenient to expand the time derivative of $Q$ in the preceding equation, considering explicitly the contribution on the normal coordinate:
\begin{eqnarray}
& &\int_s{   \frac{\mathrm{d}} {\mathrm{d}t}
 \left[  Q(\vecs{\psi}^s) \, \gamma_s( \vecs{\psi}^s) \right]\, g ^{\frac{1}{2}} \, d\vecs{\psi}^s}=\int_s{   \frac{\mathrm{d}} {\mathrm{d}t}
 \left[  Q(\vecs{\phi}^s) \, \gamma_s( \vecs{\phi}^s) \right] \, ds} = \int_s{   
 \left[   \frac{\partial (Q \gamma_s)}{\partial {\phi}_\alpha} \frac{\mathrm{d} \phi^s_\alpha} {\mathrm{d}t} +  Q \, \frac{\partial \gamma_s}{\partial t}\right]\, ds}=
 \nonumber \\
 & & \mbox{ }\int_s{   
 \left[   \frac{\partial }{ \partial {\phi}_\alpha} \left( Q \,\frac{\mathrm{d} \phi^s_\alpha} {\mathrm{d}t} \, \gamma_s \right)- Q\, \gamma_s \,\frac{\partial }{ \partial {\phi}_\alpha} \left( \frac{\mathrm{d} \phi^s_\alpha} {\mathrm{d}t} \right)  
 +  Q \, \frac{\partial \gamma_s}{\partial t}\right]\, ds}\overset{1}{=}
 \nonumber \\
 & & \mbox{ }\int_s{   
 \left[   \frac{\partial }{ \partial {\psi}^{\alpha}} \left( g ^{\frac{1}{2}} \, Q \,\frac{\mathrm{d} \psi^{s \alpha}} {\mathrm{d}t} \, \gamma_s \right)- Q \, \gamma_s \,\frac{\partial }{ \partial {\psi}^{\alpha}} \left(g ^{\frac{1}{2}} \frac{\mathrm{d} \psi^{s \alpha}} {\mathrm{d}t}\right)  +  Q \, g ^{\frac{1}{2}}\, \frac{\partial \gamma_s}{\partial t}\right]\,  d\vecs{\psi}^s }
\label{qt}
\end{eqnarray}
where equality 1 is a consequence of the expression of the divergence  $\frac{1}{g ^{\frac{1}{2}}} \frac{\partial  \left( g ^{\frac{1}{2}} \bullet^\alpha \right) }{\partial  \psi^\alpha}$  and the surface element $g ^{\frac{1}{2}} \, d\vecs{\psi}^s $ in the generalized coordinates $\vecs{\psi}$. It is reminded that $Q$ only has a implicit dependence on time which is due to the surface movement. On the other side $\gamma_s$, besides that implicit dependence, does have an explicit one, which is shown through the operator  $\frac{\partial}{\partial t}$. It should be reminded that all calculations are carried out in a Lagrangian frame of reference. In an Eulerian frame, $\gamma_s$ would have an additional implicit dependence on time through the position in physical space.

Now, it is convenient to show explicitly  the contribution of the coordinate normal to the surface ($\psi^1$ ):
\begin{eqnarray}
 & & \mbox{ }\int_s{   
 \left[   \frac{\partial }{ \partial {\psi}^{\alpha}} \left( g ^{\frac{1}{2}} \, Q \,\frac{\mathrm{d} \psi^{s \alpha}} {\mathrm{d}t} \, \gamma_s \right)- Q\,  \, \gamma_s \frac{\partial }{ \partial {\psi}^{\alpha}} \left(g ^{\frac{1}{2}} \frac{\mathrm{d} \psi^{s \alpha}} {\mathrm{d}t}\right)  \right]\,  d\vecs{\psi}^s }=
 \nonumber \\
 & & \mbox{ }\int_s{   \left[
  \frac{\partial Q}{ \partial {\psi}^{1}}  g ^{\frac{1}{2}}  \,\frac{\mathrm{d} \psi^{s 1}} {\mathrm{d}t} \, \gamma_s  - Q \, \gamma_s \,\frac{\partial }{ \partial {\psi}^{j}} \left(g ^{\frac{1}{2}} \frac{\mathrm{d} \psi^{s j}} {\mathrm{d}t}  \right)  \right]\,  d\vecs{\psi}^s },     
\label{qt2}
\end{eqnarray}
where the two terms in the normal direction have been grouped together and then $\psi^1$ derivatives expanded ($\frac{\partial \gamma_s}{ \partial {\psi}^{1}}=0$), and $$\int_s{   
 \left[   \frac{\partial }{ \partial {\psi}^{j}} \left( g ^{\frac{1}{2}} \, Q \,\frac{\mathrm{d} \psi^{s j}} {\mathrm{d}t} \, \gamma_s \right) \right]\,  d\vecs{\psi}^s }=0$$
  (the coordinate surface lines are closed, so the initial and end points of integration coincide ) has been taken into account.

Replacing  Eq. (\ref{qt2}) into Eq. (\ref{qt}) and then Eq. (\ref{qt}) into Eq. (\ref{lPDF1ext}), the following expression analogous to Eq. (\ref{lPDF1}) in generalized coordinates is obtained:
\begin{eqnarray}
 & &\frac{\mathrm{d}} {\mathrm{d}t} \int_V{Q(\vecs{\phi}) \,
{\gamma( \vecs{\phi})\, \delta_s} \, d\vecs{\phi}} =  \nonumber \\
 &   &\int_s{   
   \frac{\partial Q}{ \partial {\psi}^{1}}  \,\frac{\mathrm{d} \psi^{s 1}} {\mathrm{d}t} \, \gamma_s \, g ^{\frac{1}{2}}\, d\vecs{\psi}^s} + \int_s{ \left( \gamma_s   \frac{1}{2 g}\frac{\mathrm{d}g} {\mathrm{d}t}+   \frac{\partial \gamma_s}{\partial t}\right) \, Q\, g ^{\frac{1}{2}}\, d\vecs{\psi}^s}.
\label{lPDF1ext2}
\end{eqnarray}

On the other hand the contribution of the nonregular part analogous to Eq. (\ref{lPDF2}) is
\begin{eqnarray}
& &\int_V{
\frac{\partial Q}{\partial \phi_\alpha}   \left\langle \left.
 \frac{\mathrm{d}
c^+_\alpha}{\mathrm{d} t} \right|  \vecs{\phi} \right\rangle 
{\gamma \delta_s} \, d\vecs{\phi}} =  \int_s{
\frac{\partial Q}{\partial \phi_\alpha} \left\langle \left.
 \frac{\mathrm{d}
c^+_\alpha}{\mathrm{d} t} \right|  \vecs{\phi}^s \right\rangle
{\gamma_s } \, ds}=\nonumber \\
 & & \mbox{ } = \int_s{
\frac{\partial }{\partial \phi_\alpha} \left(  Q \left\langle \left.
 \frac{\mathrm{d}
c^+_\alpha}{\mathrm{d} t} \right|  \vecs{\phi}^s \right\rangle
{\gamma_s } \right)\, ds} 
-\int_s{
Q \frac{\partial }{\partial \phi_\alpha} \left(   \left\langle \left.
 \frac{\mathrm{d}
c^+_\alpha}{\mathrm{d} t} \right|  \vecs{\phi}^s \right\rangle
{\gamma_s } \right)\, ds}= \nonumber \\
 & & \mbox{ } = \int_s{
\frac{\partial }{\partial \psi^\alpha} \left( g ^{\frac{1}{2}} \, Q \left\langle \left.
 \frac{\mathrm{d}
c^+_\alpha}{\mathrm{d} t} \right|  \vecs{\psi}^s \right\rangle
{\gamma_s } \right)\, d\vecs{\psi}^s} -   \int_s{
Q \frac{\partial }{\partial \psi^\alpha} \left(  g ^{\frac{1}{2}} \, \left\langle \left.
 \frac{\mathrm{d}
c^+_\alpha}{\mathrm{d} t} \right|  \vecs{\psi}^s \right\rangle
{\gamma_s } \right)\, d\vecs{\psi}^s},
 \label{lPDF2ext}
\end{eqnarray}
where the same mathematical considerations as in Eq. (\ref{qt}) have been taken into account. As in Eq. (\ref{qt}),  the contribution of the coordinate normal to the surface ($\psi^1$ in Eq. (\ref{lPDF2ext}) is shown explicitly,
\begin{eqnarray}
& &\int_V{
\frac{\partial Q}{\partial \phi_\alpha}   \left\langle \left.
 \frac{\mathrm{d}
c^+_\alpha}{\mathrm{d} t} \right|  \vecs{\phi} \right\rangle 
{\gamma \delta_s} \, d\vecs{\phi}} =  \nonumber \\
 & & \mbox{ } =    \int_s{
\frac{\partial  Q}{\partial \psi^1} \,   g ^{\frac{1}{2}} \, \left\langle \left.
 \frac{\mathrm{d}
c^{1+}}{\mathrm{d} t} \right|  \vecs{\psi} \right\rangle
{\gamma_s } \, d\vecs{\psi}^s}  -   \int_s{
Q \frac{\partial }{\partial \psi^j} \left(  g ^{\frac{1}{2}} \, \left\langle \left.
 \frac{\mathrm{d}
c^{j+}}{\mathrm{d} t} \right|  \vecs{\psi} \right\rangle
{\gamma_s } \right)\, d\vecs{\psi}^s}. 
\label{lPDF3ext}
\end{eqnarray}

After  this algebra,  all the contributions are written down altogether. For this, Eq. (\ref{lPDF1ext2}) and Eq. (\ref{lPDF3ext}) are equalized, including the not-written contribution of the regular part   (Eq. (\ref{lPDF4a}) for $P^r_{\vec{c}^{+}}$ in surface coordinates). The result is :
\begin{eqnarray}
& &   \int_V{Q \, \left[
{\frac{\mathrm{d}P^r_{\vec{c}^{+}}}{\mathrm{d}t}}  +  g ^{-\frac{1}{2}}\,\frac{\partial}{\partial \psi^\alpha}\left(g ^{\frac{1}{2}}\,\left\langle
\left. \frac{\mathrm{d} c^{+\alpha}}{\mathrm{d} t} \right|
\vecs{\psi} \right\rangle P^r_{\vec{c}^{+}} \right) \right]\, g ^{\frac{1}{2}} \,d\vecs{\psi}}+
\nonumber \\
& &\int_s{ Q\,\left[ \gamma_s   \frac{1}{2 g}\frac{\mathrm{d}g} {\mathrm{d}t}+   \frac{\partial \gamma_s}{\partial t} + g ^{-\frac{1}{2}}\frac{\partial }{\partial \psi^j} \left(  g ^{\frac{1}{2}} \, \left\langle \left.
 \frac{\mathrm{d} c^{j+}}{\mathrm{d} t} \right|  {\vecs{\psi}^s} \right\rangle {\gamma_s } \right) + \right.  }\nonumber \\
 & &\mbox{   }  \left. \frac{\mathrm{d}{\psi}^1}{\mathrm{d}t}  \,
{P^r_{\vec{c}^{+}} } - \left\langle \left. \frac{\mathrm{d}
c^{1 +}}{\mathrm{d} t} \right|  {\vecs{\psi}_-^s} \right\rangle
  \, P^r_{\vec{c}^{+}}\right]\,  g ^{\frac{1}{2}}\, d\vecs{\psi}^s + \nonumber \\
& &\int_s{  \frac{\partial  Q }{ \partial {\psi}^{1}}  \, \left( \frac{\mathrm{d} \psi^{s 1}} {\mathrm{d}t} -  \left\langle \left.
 \frac{\mathrm{d}
c^{1+}}{\mathrm{d} t} \right|  {\vecs{\psi}^s} \right\rangle \right)\, \gamma_s \, g ^{\frac{1}{2}}\, d\vecs{\psi}^s} = 0 ,
\end{eqnarray}  
where $\vecs{\psi}_-^s$ indicates that the conditional average is actually the limit as the boundary is approached. Due to the existence of the singularity, the conditional temporal derivative has a discontinuity at the boundary, and this limit is the proper value to be taken.    
 
Taking into consideration that the smooth function $Q$ is arbitrary, the following equalities are obtained:
\begin{equation}
{\frac{\mathrm{d}P^r_{\vec{c}^{+}}}{\mathrm{d}t}}  +  g ^{-\frac{1}{2}}\,\frac{\partial}{\partial \psi^\alpha}\left(g ^{\frac{1}{2}}\,\left\langle
\left. \frac{\mathrm{d} c^{+\alpha}}{\mathrm{d} t} \right|
\vecs{\psi} \right\rangle P^r_{\vec{c}^{+}} \right)  = 0,
\label{trio1}
\end{equation}
\begin{equation}
 \frac{\partial \gamma_s}{\partial t} + g ^{-\frac{1}{2}}\frac{\partial }{\partial \psi^j} \left(  g ^{\frac{1}{2}} \, \left\langle \left.
 \frac{\mathrm{d} c^{j+}}{\mathrm{d} t} \right|  {\vecs{\psi}^s} \right\rangle {\gamma_s } \right) + \left(\frac{\mathrm{d}{\psi}^1}{\mathrm{d}t}  - \left\langle \left. \frac{\mathrm{d}
c^{1 +}}{\mathrm{d} t} \right|  {\vecs{\psi}_-^s} \right\rangle
  \right)  P^r_{\vec{c}^{+}}+ \gamma_s   \frac{1}{2 g}\frac{\mathrm{d}g} {\mathrm{d}t} = 0,
\label{trio2}
\end{equation}
\begin{equation} 
\frac{\mathrm{d} \psi^{s 1}} {\mathrm{d}t} \gamma_s  -  \left\langle \left. \frac{\mathrm{d} c^{1+}}{\mathrm{d} t} \right|  {\vecs{\psi}^s} \right\rangle  \, \gamma_s = 0.
\label{trio3}
\end{equation}
The meaning of these equations is clear. Eq. (\ref{trio1}) shows the known evolution of a regular PDF inside the domain $V$: once in the internal region, everything is smooth and the equation of evolution has nothing special. It corresponds to Eq. (\ref{lPDF5}) (regular PDF case). Eq. (\ref{trio2}) expresses the evolution of $\gamma_s$ (the delta weight) on the surface $s$ due to the movement of the fluid particles in the probabilistic space along the surface $s$: there is a redistribution on the surface of this weight analogous to a PDF evolution. But there is also an exchange  with $P^r_{\vec{c}^{+}}$ due to the probability current expressed by the term in $P^r_{\vec{c}^{+}}$ of the equation. If this probability current  goes inside the domain $V$, it increases the value of $P^r_{\vec{c}^{+}}$ and decreases $\gamma_s$, and the reverse in the opposite case.  Using Eqs. (\ref{trio1}) and (\ref{trio2}) in $\frac{\mathrm{d}} {\mathrm{d}t} \int_V{P^r_{\vec{c}^{+}} dV}$ shows the global balance between $P^r_{\vec{c}^{+}}$ and $\gamma_S$, which is a consequence of the normalization of the total PDF.  As  surface coordinates are used, and the surface can evolve in time, the corresponding evolution of the metrics ($g$) appear. Although a fluid, infinitely compressible, could also provide a good example, the following simple picture can help to understand the meaning of this equation: a "paperfly" is the surface with a lot of flies on it (Dirac delta). These flies can walk on the (moving) flypaper, so $\gamma_S$ evolves in time. Flies are  trapped on its surface (increase of  $\gamma_S$ at $P_{\vec{c}^{+}}$ expenses) although some of them could  escape the flypaper (decrease of  $\gamma_S$ at $P_{\vec{c}^{+}}$ benefit). The paperfly is allowed to be wrapped due to heat or some other reason (evolution of $g$). Finally, Eq. (\ref{trio3}) expresses the fact that no probability current may cross $s$, in order to keep proper normalization of the whole PDF $P_{\vec{c}^{+}}$. Notice that  this term affect only $\gamma_s$, as the singularity stays at the boundary and no jump processes exist for the fluid particles to go directly from inside to outside without passing through the boundary. Going back to the "flypaper" simple picture, the flies cannot cross the flypaper. For completing the picture, Eq. (\ref{trio1}) would represent the flies evolution in the air.

As a final remark, it is clear that the transformation of coordinates used in this appendix makes full sense in the case of spatial positions. For other quantities, the conditional values like $\left\langle \left. \frac{\mathrm{d} c^{1+}}{\mathrm{d} t} \right|  {\vecs{\psi}^s} \right\rangle$ appearing in the previous equations are better to be interpreted as ligatures in the rate of evolution of combined magnitudes.

\section{Acknowledgments}
Luis Vali\~no wants to thank Spanish Ministry of Science and Technology for its support to this work through projects PB97-1507 and BFM/2001-3320, Aragon Regional Governement through project P049/2001 and European Commission through projects HPRN-CT-1999-00041 and G14RD-CT-2000-00402.
\bibliographystyle{apsrev}
\bibliography{ZU}
\end{document}